\documentclass[prb,twocolumn,showpacs,preprintnumbers,amsmath,amssymb]{revtex4}
\usepackage{graphicx}

\begin{document}
\title{Interpolative approach for electron-electron and electron-phonon interactions: from the Kondo to the polaronic regime}
\author{A. Martin-Rodero, A. Levy Yeyati, F Flores and R. C. Monreal}
\affiliation{Departamento de F\'{i}sica Te\'{orica} de la Materia
Condensada C-V, 
Universidad Aut\'{o}noma de Madrid, E-28049 Madrid, Spain}
\begin{abstract}
We present a theoretical approach to determine the electronic properties of nanoscale systems exhibiting strong electron-electron and electron-phonon interactions and coupled to metallic electrodes. This approach is based on an interpolative ansatz for the electronic self-energy which becomes exact both in the limit of weak and strong coupling to the electrodes. The method provides a generalization of previous interpolative schemes which have been applied to the purely electronic case extensively. As a test case we consider the single level Anderson-Holstein model. The results obtained with the interpolative ansatz are in good agreement with existing data from Numerical Renormalization Group calculations. We also check our results by considering the case of the electrodes represented by a few discrete levels which can be diagonalized exactly. The approximation describes properly the transition from the Kondo regime where electron-electron interactions dominate to the polaronic case characterized by a strong electron-phonon interaction.    
\end{abstract}
\maketitle

\section{Introduction}

While Kondo physics due to electron-electron interactions have received a long standing attention \cite{Kondo-Goldhaber,Kondo-Cronenwett}, the electron phonon interaction and its interplay with electron-electron correlation effects have only attracted more recently the interest of researchers in the field of nanoscience \cite{eph1}. This is probably due to the important role played by both electron-electron and electron phonon interactions in the transport properties of small molecules \cite{eph2,eph3,eph4}. In this respect, the Anderson-Holstein model (AH) \cite{AH1,AH2,AH3,AH4,AH5}, including these correlations in a local way, is a paradigmatic simple model to study the physics of these systems.

Several theoretical methods have been successfully developed for describing Kondo correlations in the single level Anderson model like the Numerical Renormalization Group (NRG) \cite{Kondo-NRG1,Kondo-NRG2}, Quantum Monte Carlo (QMC)\cite{Kondo-QMC}, the Bethe Ansatz \cite{Kondo-BA} or perturbative methods \cite{Kondo-perturbative1,Kondo-perturbative2}. However, the extension of these methods to deal with both electron-electron and electron-phonon interactions has certain limitations. We mention here the NRG \cite{AH-NRG1,AH-NRG2,AH-NRG3,AH-NRG4} and QMC methods \cite{AH-QMC} that have been recently applied to the Anderson-Holstein Hamiltonian: in the NRG approach it is difficult to reach the strong polaronic regime in which the electronic density spectrum develops a very rich structure of phonon induced peaks over a broad range of energy; in the QMC approach, it is also difficult to resolve these features due to the limitation of using a minimum temperature in the calculations. On the other hand it would be extremely demanding from a computational point of view to implement these numerically exact methods in more complex models describing realistically systems like molecules or Quantum Dots (QD) coupled to electrodes. 
  
These limitations suggest the convenience of looking for other alternative methods that would yield approximate but sufficiently accurate solutions for these kind of systems for a broad range of parameters.
In the case of purely electronic models an approach that has proved to be very useful is the interpolative self-energy method \cite{interpolation1,interpolation2,interpolation3,interpolation4,interpolation5}. This approach is based in the analysis of the properties of the exact self-energy in the weak (perturbative) and strong coupling limits which allows to define an approximate and simple self-energy interpolating appropriately between the two limits. This scheme, originally derived for the equilibrium Anderson model, has been generalized and extensively used in different contexts: multilevel QDs \cite{interpolation-MLQD}, out of equilibrium transport through a single level \cite{interpolation-NEQD1,interpolation-NEQD2} and in combination with Dynamical Mean Field Theory (DMFT) \cite{interpolation3,interpolation4}, to analyze the Mott transition in Hubbard-like  models. Its potential to be extended to more complex situations makes this approach particularly useful.
The aim of this work is to present a generalization of this interpolative approach to deal with both electron-phonon and electron-electron interactions simultaneously.

In order to introduce the method we will consider in this paper the Anderson-Holstein model which includes both type of interactions. We first discuss the spinless case which is instructive as will allow us to illustrate how the interpolative arguments also work for the purely electron-phonon interaction case. This case will be analyzed in section II. The extension to the case where both interactions are present will be addressed in section III. In both cases the accuracy of the method is tested by comparing the results for the spectral density with exact numerical results for finite systems and with existing NRG calculations in the literature. Finally, in section IV we present a discussion and the conclusions of our work.

\section{Interpolative self-energy for the spinless Anderson-Holstein model}

We consider first the simple spinless Anderson-Holstein Hamiltonian describing a single non-degenerate electronic level, $\epsilon_0$, coupled linearly to a local phonon mode of frequency $\omega_0$ and to an electronic reservoir represented by a continuum of states of energy $\epsilon_{\vec{k}}$.
\begin{eqnarray}
\hat{H} &=& \sum_{\vec{k}}\epsilon_{\vec{k}}\hat{c}_{\vec{k}}^{\dagger}\hat{c}_{\vec{k}}+\epsilon_0\hat{c}^{\dagger}\hat{c}+\omega_{0}\hat{b}^{\dagger}\hat{b}+\lambda(\hat{b}^{\dagger}+\hat{b})\hat{n}\nonumber \\
&+&\sum_{\vec{k}}\left(V_{\vec{k}}\hat{c}_{\vec{k}}^{\dagger}\hat{c}+c.c.\right)
\label{Holstein}
\end{eqnarray} 
where $V_{\vec{k}}$ is the matrix element between the reservoir states and the localized level, $\hat{n}$ is the level occupation operator and $\lambda$ the electron-phonon coupling parameter. It should be noticed that in spite of being simpler than the spin dependent case, this model has not an exact solution for an arbitrary level occupancy. Only in the limit of a completely empty (or full) level, an exact solution is available \cite{AH3}.

As in the purely electronic case \cite{interpolation1, interpolation2} the interpolative approach will be based on a special property of the electronic self-energy which exhibits the same mathematical form when expanded in the interaction parameter both in the atomic ($V_{k}\rightarrow0$) and in the perturbative limit (which in this case corresponds to $\lambda\rightarrow0$).
  
We first consider the atomic limit, $V_{\vec{k}} \rightarrow 0$. In this case Eq. (\ref{Holstein}) can be diagonalized by means of a canonical transformation (see Refs. \cite{Holstein-canonical,Mahan}). The electron Green function can then be straightfordwardly obtained:

\begin{eqnarray}
G^{(at)}(\omega)&=&e^{-\frac{\lambda^2}{\omega_{0}^2}} 
\sum_{m=0}^{\infty}\frac{\lambda^{2m}}{\omega_{0}^{2m}m!}
\left(\frac{1-<\hat{n}>}{\omega-\tilde{\epsilon}_{0}-m\omega_{0}} 
\right. \nonumber\\
&+& \left. \frac{<\hat{n}>}{\omega-\tilde{\epsilon}_{0}+m\omega_{0}}\right)
\label{Holstein-atomic1}
\end{eqnarray}
where $\tilde{\epsilon}_0=\epsilon_0-\lambda^2/\omega_0$ and $\langle \hat{n} \rangle$ is the level occupation.

This expression can also be written as a continuous fraction which can be useful for computational purposes
\begin{eqnarray}
G^{(at)}(\omega)&=&\frac{1-<\hat{n}>}{\omega-\tilde{\epsilon}_0+\frac{\lambda^2}{\omega_0}-\frac{\lambda^2}{\omega-\tilde{\epsilon}_0+\lambda^2/\omega_0-\omega_0-\ldots}} \nonumber\\
&+& \frac{<\hat{n}>}{\omega-\tilde{\epsilon}_0-\frac{\lambda^2}{\omega_0}-\frac{\lambda^2}{\omega-\tilde{\epsilon}_0-\lambda^2/\omega_0+\omega_0-\ldots}}
\label{Holstein-atomic2}
\end{eqnarray}

The electronic self-energy in the atomic limit can now be obtained from the Dyson equation $\Sigma^{(at)}=\omega-\epsilon_H -G^{(at)-1}$ where $\epsilon_H=\epsilon_0-2(\lambda^2/\omega_0)\langle n \rangle$ is the energy level corrected by  a constant term which can be identified as the Hartree contribution. In the limit of small coupling $\lambda/\omega_0<<1$, and up to order $\lambda^2$, this expression tends to
\begin{equation}
\Sigma^{(at)}(\omega)\approx \lambda^2\left(\frac{1-<\hat{n}>}{\omega-\epsilon_0-\omega_0}+\frac{<\hat{n}>}{\omega-\epsilon_0+\omega_0}\right)
\label{Holstein-atomic3}
\end{equation}

On the other hand, the self-energy of the model can be calculated also up to order $\lambda^{2}$ by means of perturbation theory in the electron-phonon interaction. The two diagrams contributing to order $\lambda^2$ are depicted in Fig.1a) and Fig.1b). Diagram a) gives simply the constant Hartree contribution $-2(\lambda^2/\omega_0){\langle n \rangle}_0$, where ${\langle n \rangle}_0$ is the unperturbed level occupation for an effective level , $\epsilon_{eff}$, which will be fixed by an appropriate self-consistency condition as commented below.
\begin{figure}
\includegraphics[width=0.7\columnwidth]{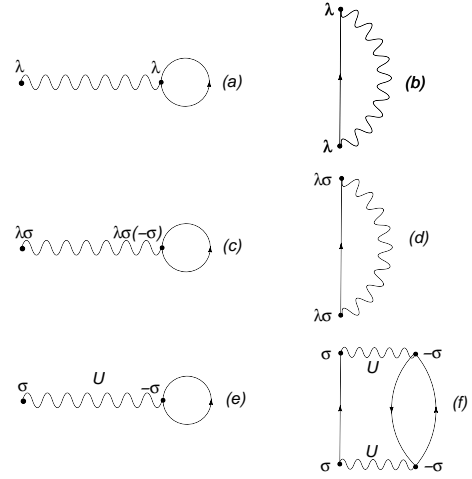}
\caption{Lower order diagrams contributing to the proper self-energy. Diagrams a) and b) correspond to the spinless case. Diagrams c), d), e) and f) to the spin dependent Anderson-Holstein model.}    
\label{diagrams}
\end{figure}

The correlated part of the self-energy to order $\lambda^2$ is then given by the diagram of Fig.1b and has the expression:
\begin{eqnarray}
\Sigma^{(2)}(\omega)&=&\lambda^2\left(\int_{\mu}^{\infty}d\epsilon\frac{\rho^{(0)}(\epsilon)}{\omega-\epsilon-\omega_0+i\eta} \right. \nonumber\\
&&+ \left. \int_{-\infty}^{\mu}d\epsilon\frac{\rho^{(0)}(\epsilon)}{\omega-\epsilon+\omega_0+i\eta}\right)
\label{Holstein-self2}
\end{eqnarray}
where $\rho^{(0)}(\omega)=\Gamma/\left[(\omega-\epsilon_{eff})^2+\Gamma^2\right]/\pi$ is the level density of states of the one-electron unperturbed case, $\Gamma=\pi\sum_k|V_k|^2\delta(\omega-\epsilon_k)$ which is taken as constant and $\mu$ is the reservoir chemical potential.

In the limit $\Gamma\rightarrow0$ the above expression tends to
\begin{equation}
\Sigma^{(2)}(\omega)\rightarrow\lambda^2\left(\frac{1-<\hat{n}>_0}{\omega-\epsilon_{eff}-\omega_0}+\frac{<\hat{n}>_0}{\omega-\epsilon_{eff}+\omega_0}\right)\equiv F(\omega)
\label{Holstein-self2-atomic}
\end{equation}

By comparing Eqs. (\ref{Holstein-atomic3}) and (\ref{Holstein-self2-atomic}) one can notice that the exact atomic self-energy up to order $\lambda^2$ and the second order self-energy in the limit $\Gamma\rightarrow0$ have the same functional dependence on the occupations and frequency, as was the case in the purely electronic case. Therefore it is possible to find an interpolation following the lines discussed in Refs. \cite{interpolation1,interpolation2}, the only difference being the more complex expression for the atomic self-energy due to the multi-phonon satellite structure. The interpolative self-energy is defined by the following ansatz: 

\begin{equation}
\Sigma(\omega)=\Sigma^{(at)}\left\{F^{-1}\left[\Sigma^{(2)}(\omega)\right]\right\}
\label{interpol1}
\end{equation}
where $F^{-1}$ is the inverse function defined by Eq. (\ref{Holstein-self2-atomic}).  
It is straightforward to check that this ansatz interpolates correctly between the weak and strong coupling limits. The atomic limit is recovered because in the limit $\Gamma/\lambda\rightarrow0$, $F^{-1}\left[\Sigma^{(2)}(\omega)\right]\rightarrow\omega$. On the other hand, in the limit of small electron-phonon coupling, $\Sigma^{(at)}\left\{F^{-1}\left[\Sigma^{(2)}(\omega)\right]\right\}\rightarrow\Sigma^{(2)}(\omega)$, and the results of perturbation theory are recovered.
>From Eq. (\ref{Holstein-self2-atomic}), we can straightfordwardly calculate $F^{-1}\left[\Sigma^{(2)}(\omega)\right]\equiv\Omega(\omega)$ which is given by the roots of a quadratic equation:
\begin{widetext}
\begin{equation}
\Omega(\omega)=\epsilon_{eff}+\frac{\lambda^2}{2\Sigma^{(2)}}\left[1+\sqrt{1+\frac{4\omega_0\Sigma^{(2)}}{\lambda^2}(1-2<n>_0)+  \left( \frac{2\omega_0{\Sigma^{(2)}}}{\lambda^2} \right)^2} \right]
\label{quadratic}
\end{equation}
\end{widetext}
where the $+$ sign in front of the square root has been chosen to yield $\Omega(\omega\rightarrow\pm\infty)= \omega$; this guarantees the recovery of the atomic limit for $\Sigma(\omega)$ if $\Gamma/\lambda \rightarrow 0$. Moreover, we have checked that  Eq. (\ref{quadratic}) defines an analytical function of $\omega$ with poles in the positive complex half-plane: this is a necessary condition to have a self-energy with the correct analytical properties.

We now discuss how to determine $\epsilon_{eff}$ by imposing an appropriate self-consistency condition. Several procedures have been proposed in the literature: The simplest one is to impose charge consistency \cite{interpolation1,interpolation2,interpolation-NEQD1} between the one-electron and interacting problems, $\langle n \rangle={\langle n \rangle}_0$. As discussed in \cite{interpolation-NEQD1} this procedure only fulfills the Friedel sum-rule in an approximate way although the agreement is rather good for a broad range of parameters. An alternative possibility is to fix the effective one-electron level by imposing the fulfillment of the Friedel sum rule \cite{interpolation3}. The self-consistency condition is then given by the equation:
\begin{equation}
<n> = \frac{1}{2}-\frac{1}{\pi}\arctan\left(\frac{\epsilon_H+\Sigma(0)}{\Gamma}\right)
\label{Friedel}
\end{equation} 
where the occupation $\langle n \rangle$ is calculated from the interacting retarded propagator $G^r(\omega)=[\omega-\epsilon_H-\Sigma(\omega)]^{-1}$:
\begin{equation}
<n> = -\frac{1}{\pi}\int^{\mu}_{-\infty}d\omega \mbox{Im} G^{r}(\omega)
\label{occupation}
\end{equation}  
This last condition seems to give a better description of the density of states around the Fermi level, specially in cases with a large electron-hole asymmetry.

\begin{figure}[b!]
\includegraphics[width=0.8\columnwidth]{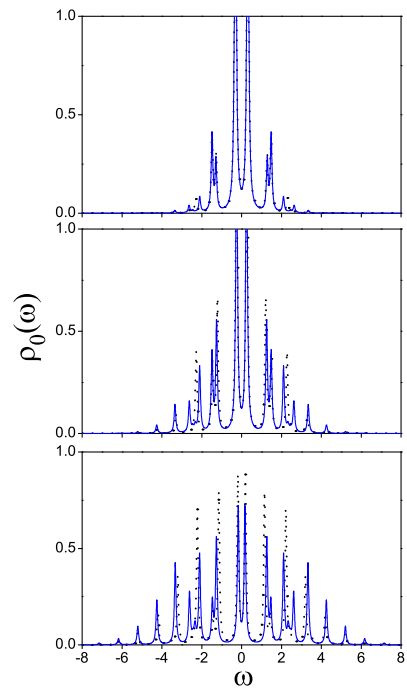}
\caption{(Color online) Localized level spectral density for a discrete spinless Anderson-Holstein model with an electrode consisting of a three atom  tight-binding chain for different values of the electron-phonon coupling parameter $\lambda$ with $\omega_0=1$. From upper to lower panel, $\lambda=0.6,1,1.4$. The (blue) continuous line corresponds to the interpolative solution while the (black) dotted one to the numerical diagonalization results. The diagonal energy levels in the chain are taken as zero and the first-neighbors hopping parameter is fixed at $t=1$. The hopping parameter between the dot level and the chain is $V=0.5$ and the level position is the one corresponding to an e-h symmetric case. A small imaginary part, $\eta=0.05$, is included in both spectral densities to facilitate the comparison.} 
\label{spinless-cluster}
\end{figure}

This interpolative ansatz reproduces very accurately both the limit of weak and strong coupling regimes while giving a good quantitative description in the intermediate range. In order to check this approximation we first compare its predictions for a discrete version of the spinless Anderson-Holstein Hamiltonian in which the electrode is replaced by a three site tight-binding chain which can be numerically diagonalized. 
This is a useful model which allows a detailed comparison with exact results between the weak and strong coupling regimes.  For a given value of the electron-phonon coupling $\lambda$, the configuration Hilbert space is truncated in the number of phonon modes in such a way that convergence in the level spectral density is achieved. 

In Fig. \ref{spinless-cluster} we compare the level spectral density, $\rho_0(\omega)=-\mbox{Im} G^r(\omega)/\pi$, obtained from the numerical diagonalization with the approximate solution given by the interpolative self-energy for a case with electron-hole (e-h) symmetry.
As shown in the figure, the interpolative approximation yields in all the cases an excellent approximation to the exact DOS. Notice that the range of parameters goes from the weak coupling limit ($\lambda=0.6$), to the intermediate ($\lambda=1$) and the strong coupling case ($\lambda=1.4$). For lower and higher values of $\lambda$ both spectra are practically indistinguishable. 

\begin{figure}[b]
\includegraphics[width=1.0\columnwidth]{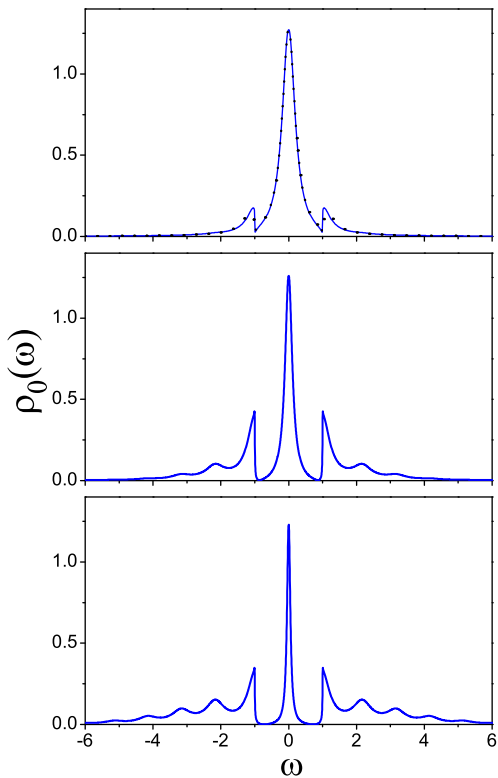}
\caption{(Color online) Localized level spectral density for the continuous spinless Anderson-Holstein model for different values of the electron-phonon coupling parameter for an e-h symmetric case with $\Gamma=0.25$ and $\omega_0=1$ (from upper to lower panel, $\lambda=0.3,1,1.4$). The (blue) continuous line corresponds to the interpolative solution and the (black) dotted line in the upper panel to the second-order self-energy approximation.} 
\label{spinless-continuum}
\end{figure}
Results for the continuous model corresponding to Hamiltonian Eq. (\ref{Holstein}) are shown in Fig. \ref{spinless-continuum}.
As in Fig. \ref{spinless-cluster} the three cases go from the weak to the strong coupling limit showing the interpolative solution and the results provided by second-order perturbation theory in the self-energy. It is interesting to realize that our interpolative solution is a fair approximation of the second order perturbation theory for small $\lambda$; however, for the intermediate or the strong coupling limit, our solution develops the typical structure of the multiphonon excitation spectrum which is more clearly seen at high $\lambda$. In addition to this multi-phonon features, the DOS exhibits a narrowing resonance at the Fermi level when approaching the strong coupling limit. It is straightforward to calculate the weight of this resonance from the value of the self-energy at the Fermi level, giving an exponential decaying law $\approx exp(-\lambda^2/\omega_0^2)$.

\begin{figure}
\includegraphics[width=1.0\columnwidth]{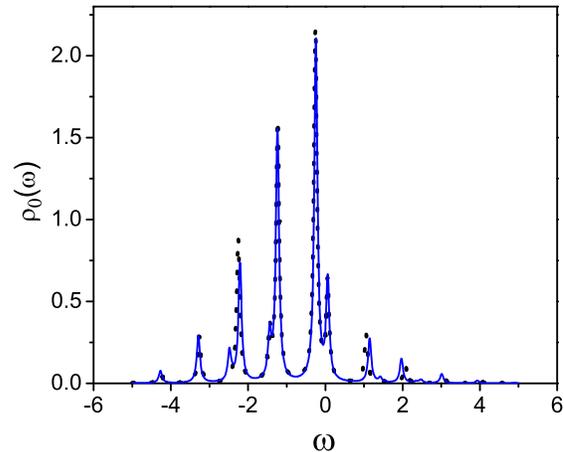}
\caption{(Color online) Localized level spectral density for a discrete spinless Anderson-Holstein model with an electrode consisting of a three atom  tight-binding chain for an asymmetric electron-hole case. The (blue) continuous line corresponds to the interpolative solution while the (black) dotted one to the numerical diagonalization results. The renormalized level position is $\tilde{\epsilon_0}=-0.2$ with $\lambda=1$ and $\omega_0=1$. The diagonal energy levels in the chain are taken as zero and the first-neighbors hopping parameter in the chain is fixed at $t=1$. The hopping parameter between the dot level and the chain is $V=0.25$. A small imaginary part, $\eta=0.05$, is included in both spectral densities to facilitate the comparison.} 
\label{spinless-cluster-asim}
\end{figure}

\begin{figure}
\includegraphics[width=1.0\columnwidth]{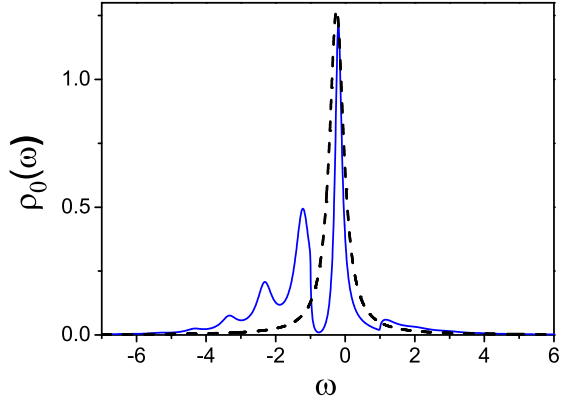}
\caption{(Color online) Localized level spectral density for the continuous spinless Anderson-Holstein model for an electron-hole asymmetric case. The parameters values are $\lambda=1$, $\Gamma=0.25$, $\omega_0=1$ and $\tilde{\epsilon}_{0}=-0.25$. The (blue) continuous line corresponds to the interpolative solution and the (black) dashed line to the non-interacting case with $\epsilon_0=\tilde{\epsilon}_0$.} 
\label{spinless-continuum-asim}
\end{figure}

For the sake of completeness we show in Figs. \ref{spinless-cluster-asim} and \ref{spinless-continuum-asim} results for the level DOS for an electron-hole asymmetric case. In Fig. \ref{spinless-cluster-asim} we compare our interpolative solution with exact diagonalization results for the same discrete model used in Fig. \ref{spinless-cluster}. Notice the excellent agreement between both solutions which correspond to a level occupancy $\langle n_{diag} \rangle \approx \langle n \rangle =0.83$. In Fig. \ref{spinless-continuum-asim} we show the local DOS for the full continuous asymmetric model for a case with a level occupancy $\langle n \rangle =0.82$ (see caption for details). Notice how the non-interacting resonance (blue dashed line) having a Lorentzian-like shape around $\tilde{\epsilon}_0$ is changed into a density of states showing a satellite structure associated with phonons.      

As a final remark, we would like to point out that the interpolative ansatz recovers the exact solution in the limit of a fully empty or fully occupied level. In this case $\Omega\approx\omega+i\Gamma$ and the exact propagator 
in this limit, $G(\omega)=G^{(at)}(\omega + i \Gamma)$ is recovered \cite{AH3}. 

\section{The spin degenerate Anderson-Holstein Hamiltonian}
In this section we will show how this interpolative scheme can be extended to the presence of both electron-electron and electron-phonon interactions, by considering the spin degenerate Anderson-Holstein Hamiltonian:

\begin{eqnarray}
\hat{H} &=& \sum_{\vec{k}\sigma}\epsilon_{\vec{k}}\hat{n}_{\vec{k}}+\epsilon_0\sum_{\sigma}\hat{c}_{\sigma}^{\dagger}\hat{c}_{\sigma}+U\hat{n}_{\uparrow}\hat{n}_{\downarrow}+\omega_{0}\hat{b}^{\dagger}\hat{b} \nonumber\\
&+&\lambda(\hat{b}^{\dagger}+\hat{b})\sum_{\sigma}\hat{n}_{\sigma}+\sum_{\vec{k}\sigma}\left(V_{\vec{k}}\hat{c}_{\vec{k}\sigma}^{\dagger}\hat{c}_{\sigma}+c.c.\right)
\label{Anderson-Holstein}
\end{eqnarray}
where $U$ is the Coulomb interaction in the localized level.
The atomic Green function can also be obtained from a canonical transformation of this Hamiltonian in the limit $V_k\rightarrow0$ giving  a renormalization of both the level position $\tilde{\epsilon}_0=\epsilon_0-\lambda^2/\omega_0$, and the Coulomb interaction $\tilde{U}=U-2\lambda^2/\omega_0$. As a function of these parameters the atomic Green function takes the form \cite{AH-NRG1}: 
\begin{widetext}
\begin{eqnarray}
G_{\sigma}^{(at)}(\omega)&=&e^{-\frac{\lambda^2}{\omega_{0}^2}} \sum_{m=0}^{\infty}\frac{\lambda^{2m}}{\omega_{0}^{2m}m!} \left( \frac{<(1-\hat{n}_{\sigma})(1-\hat{n}_{\bar{\sigma}})>}{\omega-\tilde{\epsilon}_{0}-m\omega_{0}}+\frac{<(1-\hat{n}_{\sigma})\hat{n}_{\bar{\sigma}}>}{\omega-\tilde{\epsilon}_{0}-\tilde{U}-m\omega_{0}} \right. \nonumber\\
&+& \left. \frac{<\hat{n}_{\sigma}(1-\hat{n}_{\bar{\sigma}})>}{\omega-\tilde{\epsilon}_{0}+m\omega_{0}}+ \frac{<\hat{n}_{\sigma}\hat{n}_{\bar{\sigma}}>}{\omega-\tilde{\epsilon}_{0}-\tilde{U}+m\omega_{0}} \right)
\label{AH-atomic1}
\end{eqnarray}
\end{widetext}
For computational purposes, this expression can also be written in a continuous fraction form.
Following an analogous procedure to the one used in the spinless case, the atomic self-energy can be calculated from the above Green function as $\Sigma^{(at)}=\omega-\epsilon_{H\sigma}-G^{(at)-1}$ where the Hartree level has in this case the expression
\begin{equation}
\epsilon_{H\sigma}=\epsilon_0-\frac{2\lambda^2}{\omega_0}\sum_{\sigma}<\hat{n}_{\sigma}>+U<\hat{n}_{\bar{\sigma}}>
\end{equation}
Expanding now the correlated part of the atomic self-energy up to order $\lambda^2$ and $U^2$ we obtain the equation:
\begin{eqnarray}
\Sigma_{\sigma}^{(at)}(\omega)&\approx& \lambda^2\left(\frac{1-<\hat{n}_{\sigma}>}{\omega-\epsilon_0-\omega_0}+\frac{<\hat{n}_{\sigma}>}{\omega-\epsilon_0+\omega_0}\right)\nonumber\\
&+&U^2\frac{<\hat{n}_{\bar{\sigma}}>(1-<\hat{n}_{\bar{\sigma}}>)}{\omega-\epsilon_0}
\label{AH-atomic2}
\end{eqnarray}
On the other hand, the self-energy calculated by means of the lowest order diagrams (see Fig. 1d) and f))has the following expression in the $\Gamma\rightarrow0$ limit (up to the Hartree constant contribution given by diagrams 2c) and 2e)):

\begin{eqnarray}
\Sigma^{(2)}_\sigma(\omega)&\rightarrow& \lambda^2\left(\frac{1-<\hat{n}_{\sigma}>_0}{\omega-\epsilon_{eff}-\omega_0}+\frac{<\hat{n}_{\sigma}>_0}{\omega-\epsilon_{eff}+\omega_0}\right)\nonumber\\
&+& U^{2}\frac{<\hat{n}_{\bar{\sigma}}>_0(1-<\hat{n}_{\bar{\sigma}}>_0)}{\omega-\epsilon_{eff}}\equiv F_{\sigma}(\omega)
\label{AH-self2}
\end{eqnarray}

Notice that when both electron-electron and electron-phonon interactions are present, both equations have still the same functional dependence. Thus one can introduce an interpolative self-energy using the same approach developed for the spinless case; in particular we define the self-energy by the \textit{ansatz}:
\begin{equation}
\Sigma_{\sigma}(\omega)=\Sigma^{(at)}_{\sigma}\left\{F_{\sigma}^{-1}\left[\Sigma^{(2)}_{\sigma}(\omega)\right]\right\}
\label{AH-ansatz}
\end{equation}
$F_{\sigma}^{-1}$ being the inverse function defined by Eq. (\ref{AH-self2}). These equations define the interpolative self-energy we are looking for, yielding the appropriate behavior for small values of $\lambda$ and $U$, and in the atomic limit, when $V_{k}\rightarrow0$. An important difference with the spinless case appears because the atomic selfenergy for the Anderson-Holstein Hamiltonian depends on the correlation function $\langle \hat{n}_\sigma\hat{n}_{\bar{\sigma}} \rangle$ which has to be calculated in a self-consistent way. This is discussed in the appendix.

In order to obtain $F_{\sigma}^{-1}\left[\Sigma^{(2)}_{\sigma}(\omega)\right]\equiv\Omega_{\sigma}(\omega)$ from Eq. (\ref{AH-self2}) it is now necessary to solve the following cubic equation:
\begin{eqnarray}
&&\tilde{\Omega}_{\sigma} (\tilde{\Omega}_{\sigma}^2-\omega_0^2) \Sigma^{(2)}_{\sigma}=\lambda^2 \tilde{\Omega}_{\sigma}\left[\tilde{\Omega}_{\sigma}+(1-2<n_{\sigma}>_0)\:\omega_0\right]\nonumber\\
&+& U^2<n_{\bar{\sigma}}>_0(1-<n_{\bar{\sigma}}>_0)(\tilde{\Omega}_{\sigma}^2-\omega_0^2)
\label{cubic}
\end{eqnarray}
where $\tilde{\Omega}_{\sigma}=\Omega_{\sigma}-\epsilon_{eff}$.
This equation has two clear limiting cases. In the limit $\tilde{U}>>0$ and $U>>\lambda$, the solution of Eq. (\ref{cubic}) tends to
\begin{equation}
\Omega_{\sigma}(\omega)=\epsilon_{eff}+\frac{U^2<n_{\bar{\sigma}}>_0(1-<n_{\bar{\sigma}}>_0)}{\Sigma^{(2)}_{\sigma}(\omega)}
\label{electronic}
\end{equation}
which corresponds to the purely electronic case in the absence of electron-phonon interaction (see Refs. \cite{interpolation1,interpolation2,interpolation3}). On the other hand, in the opposite limit, $\tilde{U}<<0$ and $U<<\lambda$, Eq. (\ref{cubic}) reduces to a quadratic equation like the one in the spinless case and with a similar solution (Eq. (\ref{quadratic})) which would correspond to the strong polaronic limit: 
\begin{equation}
(\tilde{\Omega}_{\sigma}^2-\omega_0^2) \Sigma^{(2)}_{\sigma}=\lambda^2 \left[\tilde{\Omega}_{\sigma}+(1-2<n_{\sigma}>_0)\:\omega_0\right]
\label{polaronic}
\end{equation}

Solving the full cubic equation requires choosing the root that has the correct physical behavior. As in the spinless case this can be achieved by starting from the solution that behaves as $\Omega(\omega\rightarrow \pm\infty)\approx \omega$ and imposing continuity for all frequencies. This solution exhibits a transition from a regime with $\tilde{U}>0$ in which $\Omega_{\sigma}(\omega)$ qualitatively behaves like the one in the electronic case to a regime for $\tilde{U}<0$ in which $\Omega_{\sigma}(\omega)$ rather behaves like the one corresponding to the quadratic equation of the polaronic limit.
The transition takes place around $\tilde{U}\approx0$, its precise value as a function of $\lambda$, $U$ and $\Gamma$ can be calculated by a careful analysis of the analytical properties of the roots of Eq. (\ref{cubic}). 
We have furthermore checked that in the polaronic regime ($\tilde{U}<<0$), the solutions of Eq. (\ref{cubic}) are well described by the roots of the much more simple quadratic equation:
\begin{eqnarray}
&&(\tilde{\Omega}_{\sigma}^2-\omega_0^2) \Sigma^{(2)}_{\sigma}=\lambda^2 \left[\tilde{\Omega}_{\sigma}+(1-2<n_{\sigma}>_0)\:\omega_0\right]\nonumber\\
&+&U^2<n_{\bar{\sigma}}>_0(1-<n_{\bar{\sigma}}>_0)\tilde{\Omega}_{\sigma}
\label{polaronic2}
\end{eqnarray}
which is obtained by neglecting in Eq. (\ref{cubic}) the term $-U^2<n_{\bar{\sigma}}>_0(1-<n_{\bar{\sigma}}>_0)\omega_0^2$. It is easy to check that both limits $\omega_0\rightarrow0$ and $\lambda/U\rightarrow\infty$ are recovered by this simpler equation.

\begin{figure}
\includegraphics[width=1.0\columnwidth]{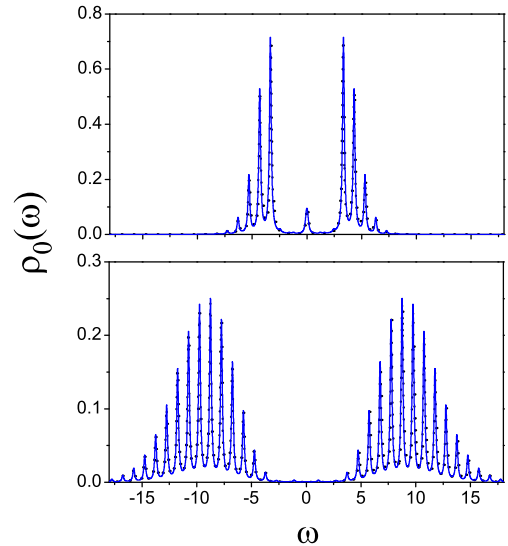}
\caption{(Color online) Localized level spectral density for a discrete Anderson-Holstein model with the electrode simulated by a single discrete level (taken as the zero of energy) for different values of the electron-phonon coupling $\lambda$ with $\omega_0=1$. From upper to lower panel, $\lambda=0.9,2.6$. The (blue) continuous line corresponds to the interpolative solution while the (black) dotted one to the numerical diagonalization results. The hopping parameter between the dot level and the chain is $V=0.25$ and the dot level position is the one corresponding to an e-h symmetric case. A small imaginary part, $\eta=0.1$, is included in both spectral densities to facilitate the comparison. The Coulomb interaction parameter is $U=8$.} 
\label{spin-cluster}
\end{figure}

In order to test the interpolative solution presented in this section we first compare its predictions with the results of exact numerical diagonalizations for a finite system. As the Hilbert space in the case were both electron-electron and electron-phonon interactions are present is much larger than in the spinless case we take a minimum cluster in which the electrode is replaced by a single discrete level. In Fig. \ref{spin-cluster} we represent the evolution of the level DOS for increasing values of the electron-phonon coupling parameter $\lambda$. It is interesting to notice that in spite of the simplicity of this discrete model the local DOS exhibits as a function of $\lambda$ a transition from a regime where Coulomb correlations dominate (Kondo regime, see upper panel) to a strong polaronic regime where the effective electron-electron interaction becomes negative and is characterized by a gap in the quasi-particle spectrum and the appearance of well defined multi-phonon satellites (see lower panel). As can be observed the interpolative solution is very close to the exact diagonalization results throughout these regimes.

\begin{figure}
\includegraphics[width=1.0\columnwidth]{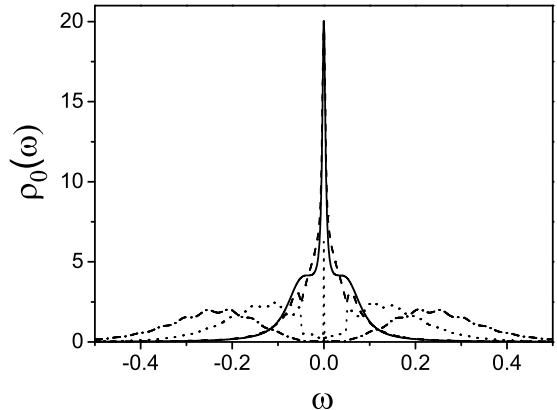}
\caption{Localized level spectral density of the continuous Anderson Holstein model for a fixed value of the Coulomb interaction $U$ and for different values of the electron-phonon coupling constant. The magnitude of the parameters has been chosen as to allow a comparison with existing NRG calculations . $\Gamma=0.0159$, $\omega_0=0.05$, $U=0.1$ and $\lambda=0,0.03,0.07,0.09$ (corresponding respectively to continuous, dashed, dotted and dash-dotted lines). The level position corresponds to an electron-hole symmetric case.} 
\label{spin-continuum}
\end{figure}

In Fig. \ref{spin-continuum} we show our calculations for the localized level spectral density for the full Anderson-Holstein Hamiltonian of Eq. (\ref{Anderson-Holstein}). The parameters values have been chosen to allow comparison with NRG results for an electron-hole symmetric case \cite{AH-NRG1,AH-NRG2} and illustrate the transition from the Kondo to the strong polaronic regime. For $\lambda=0$ we find the typical spectrum with a Kondo resonance at the Fermi level, in good agreement with \cite{AH-NRG1}. When increasing $\lambda$, the effective Coulomb repulsion $\tilde{U}$ decreases and the Fermi-level resonance slightly broadens while the first phonon sub-bands appear (see curve for $\lambda=0.03$ and $\tilde{U}=0.064$ in the figure). For larger values of $\lambda$, $\tilde{U}$ becomes negative, the Fermi level resonance eventually collapses, and the system experiments a transition to a polaronic regime with an energy gap in the spectral density. Strictly speaking, the Friedel sum rule is still satisfied, with a delta-like peak at the Fermi energy which at the scale of the figure cannot be resolved. These results are in good agreement with those of NRG calculations \cite{AH-NRG1,AH-NRG2}.  

The results presented so far in this section have concentrated in the case of electron-hole symmetry. The interpolative ansatz is however valid as in the spinless case for a general asymmetric situation. In the extreme limit of a fully empty or fully occupied level, the model becomes in fact equivalent to a spinless case as it is evident comparing Eq. (\ref{AH-atomic1}) with Eq. (\ref{Holstein-atomic1}) for $\langle \hat{n}_{\sigma} \rangle=0$, $(\langle \hat{n}_{\sigma}\hat{n}_{\bar{\sigma}} \rangle=0)$ or $\langle \hat{n}_{\sigma} \rangle=1$, $(\langle \hat{n}_{\sigma}\hat{n}_{\bar{\sigma}} \rangle=1)$. Therefore the ansatz tends also to the exact solution in that limit.  
 
\section{Conclusions}
We have presented an interpolative approach for calculating the electronic properties in a nanoscale system coupled to metallic electrodes and exhibiting both electron-electron and electron-phonon interactions taking the Anderson-Holstein model as a test case. This approach provides a generalization of the interpolative self-energy method \cite{interpolation1,interpolation2,interpolation3} which has been applied successfully to many different problems \cite{interpolation3,interpolation4,interpolation-MLQD,interpolation-NEQD1,interpolation-NEQD2}. 

In a first step we have considered the spinless Anderson-Holstein model and discussed in detail for this case how the interpolative argument can be extended to deal with electron-phonon interactions. We have checked the accuracy of the approximation by applying the method to the case in which the electrode is represented by a finite chain of three discrete levels which can be solved exactly by numerical diagonalization. In particular we have found that the method gives a very good description of the spectral density at low energies. In the continuous case the method describes correctly the exponential decay of the width of the Fermi level resonance with increasing electron-phonon coupling.

In a second step we have analyzed the full spin dependent Anderson-Holstein model and discussed how to generalize the interpolative self-energy for this case. This scheme naturally recovers the results of the purely electronic interpolation method for vanishing electron-phonon coupling. The accuracy of the method for this case is checked both for a discrete electrode model and by comparison with existing NRG results for the continuous model \cite{AH-NRG1,AH-NRG2}. 

We conclude that the interpolative technique is very reliable and provides a powerful tool to explore the role of strong electron-phonon interactions in more general situations of current experimental interest like electronic transport through single molecules. Work along these lines is under progress. 

\begin{acknowledgements}      
We thankl L. Arrachea for fruitful discussions. Support by Spanish CYCIT through contracts FIS2005-06255 and
FIS2005-02909 is acknowledge.
\end{acknowledgements}
\appendix
\section{}

For a purely electronic case the correlation function $\langle \hat{n}_{\uparrow}\hat{n}_{\downarrow} \rangle$ can be calculated from the one-electron Green function of the localized level \cite{interpolation-MLQD}.
A similar procedure for the Anderson-Holstein model yields the following equation: 

\begin{equation}
U<\hat{n}_{\sigma}\hat{n}_{\bar{\sigma}}>+\lambda<\hat{n}_{\sigma}(\hat{b}^{\dagger}+\hat{b})> = \frac{1}{2\pi i} \oint d\omega\ \Sigma_{\sigma}(\omega)G_{\sigma}(\omega)
\end{equation}
where the propagators are assumed to be causal and the contour integration is on the upper complex plane.

This equation could be used to calculate $\langle \hat{n}_{\sigma}\hat{n}_{\bar{\sigma}} \rangle$ if the averaged value $\langle \hat{n}_{\sigma}(\hat{b}^{\dagger}+\hat{b}) \rangle$ can be obtained independently. This can be provided by the equation of motion of the phonon propagator.

\begin{equation}
D(\omega)=D^{(0)}(\omega)+\lambda D^{(0)}(\omega)<<\sum_\sigma \hat{n}_{\sigma};\hat{b}^{\dagger}+\hat{b}>>_\omega \nonumber 
\end{equation}
where the Green function appearing on the right hand side can be used to calculate the average $\langle \hat{n}_{\sigma}(\hat{b}^{\dagger}+\hat{b}) \rangle$. From the equation of motion of this last Green function we have \cite{AH-NRG1}:
\begin{equation}
<<\sum_{\sigma}\hat{n}_{\sigma};\hat{b}^{\dagger}+\hat{b }>>_{\omega}=\lambda D^{(0)}(\omega)<<\sum_{\sigma}\hat{n}_{\sigma};\sum_{\sigma}\hat{n}_{\sigma}>>_{\omega} 
\label{equation-of-motion}
\end{equation}

The average value $\langle \sum_{\sigma}\hat{n}_{\sigma}(\hat{b}^{\dagger}+\hat{b}) \rangle$ can be calculated by integrating the above equation:
\begin{equation}
<\sum_{\sigma}\hat{n}_{\sigma}(\hat{b}^{\dagger}+\hat{b })>=\lambda\oint\frac{d\omega}{2\pi i}D^{(0)}(\omega)\chi_{Q}(\omega)
\label{charge-correlation}
\end{equation}
where $\chi_{Q}(\omega)$ is the charge susceptibility appearing in Eq. \ref{equation-of-motion}.

An interpolative argument can also be used to calculate $\chi_{Q}(\omega)$. First, we consider the atomic limit ($\Gamma\rightarrow0$) and use the canonical transformation that diagonalizes the Anderson-Holstein Hamiltonian in this case. We obtain:
\begin{equation}
\chi_{Q}^{(at)}(\omega)=-2\pi i\sum_{\sigma}\left[<\hat{n}_{\sigma}>+<\hat{n}_{\sigma}\hat{n}_{\bar{\sigma}}> \right] \delta(\omega)
\label{charge-correlation-atomic}
\end{equation}

On the other hand, in the limit $\lambda, U\rightarrow0$ we can evaluate this charge susceptibility function using Wick's theorem:
\begin{eqnarray}
\chi_{Q}^{(Wick)}(\omega)&=&-2\pi i\sum_{\sigma}\left[<\hat{n}_{\sigma}>^2+<\hat{n}_{\sigma}><\hat{n}_{\bar{\sigma}}> \right] \delta(\omega)\nonumber\\
&+&\sum_{\sigma}\chi_{\sigma}^{(0)}(\omega)
\label{charge-correlation-Wick}
\end{eqnarray}
where $\chi_{\sigma}^{(0)}(\omega)$ is the lowest order polarization bubble. In the limit $\Gamma\rightarrow0$, $\chi_{\sigma}^{(0)}(\omega)\rightarrow -2\pi i \langle \hat{n}_{\sigma} \rangle (1-\langle \hat{n}_{\sigma} \rangle ) \delta(\omega)$ and Eq. \ref{charge-correlation-Wick} tends to:

\begin{equation}
\chi_{Q}^{(Wick)}(\omega)=-2\pi i\sum_{\sigma}\left[<\hat{n}_{\sigma}>+<\hat{n}_{\sigma}><\hat{n}_{\bar{\sigma}}> \right] \delta(\omega)
\label{charge-correlation-Wick2}
\end{equation}

Due to these properties, the function:
\begin{equation}
\chi_{Q}(\omega)=-2\pi i\sum_{\sigma}\left[<\hat{n}_{\sigma}>^2+<\hat{n}_{\sigma}\hat{n}_{\bar{\sigma}}> \right] \delta(\omega)+\sum_{\sigma}\chi_{\sigma}(\omega)
\label{charge-correlation-interpolative}
\end{equation}
interpolates correctly between both limits, $\chi_{\sigma}(\omega)$ being the full polarization bubble. We have checked that a simple effective mass approximation in which $\chi_{\sigma}(\omega)$ is calculated by
\begin{equation}
\chi_{\sigma}(\omega)=\int d\tau G_{\sigma}^{eff}(\tau)G_{\sigma}^{eff}(-\tau)e^{-i\omega\tau}
\label{effective-bubble}
\end{equation}
where $G_{\sigma}^{eff}$ is a renormalized propagator with $V^{eff}=V/(1-\partial\Sigma(\mu)/\partial\omega)^{1/2}$, gives a good estimation of $\langle \hat{n}_{\sigma}\hat{n}_{\bar{\sigma}} \rangle$ for a broad range of parameters. As an additional check  $\langle \hat{n}_{\sigma}\hat{n}_{\bar{\sigma}} \rangle$ has been obtained by means of clusters calculations with a good agreement with the above approximation.

\end{document}